\documentclass[twocolumn, aps, prl, 10pt]{revtex4}
\usepackage{graphicx}
\usepackage{lipsum}
\usepackage{braket}
\usepackage{amsmath}
\usepackage{xfrac}
\usepackage{natbib}
\usepackage{xcolor}
\usepackage{array}

\graphicspath{{./Figures/}}

\begin{document}

\title{Coherent metamaterial absorption of two-photon states with 40\% efficiency.}

\author{Ashley Lyons$^{1,2,*}$, Dikla Oren$^{3,*}$, Thomas Roger$^{2}$, Vassili Savinov$^{4}$, Jo\~{a}o Valente$^{4,5}$, Stefano Vezzoli$^{2}$, Nikolay I. Zheludev$^{4}$, Mordechai Segev$^{3}$, Daniele Faccio$^{1,2}$}

\affiliation{$^{1}$School of Physics and Astronomy, University of Glasgow, Glasgow, G12 8QQ, UK\\
$^{2}$School of Engineering and Physical Sciences, Heriot-Watt University, Edinburgh, EH14 4AS, UK\\ 
$^{3}$Physics Department and Solid State Institute, Technion, 32000 Haifa, Israel\\
$^{4}$Optoelectronics Research Centre \& Centre for Photonic Metamaterials, University of Southampton, Southampton SO17 1BJ, UK.\\
$^{5}$Department of Electronic \& Electrical Engineering, University College London, London, WC1E 7JE
}

\date{\today}

\begin{abstract}{Multi-photon absorption processes have a nonlinear dependence on the amplitude of the incident optical field i.e. the number of photons. However, multi-photon absorption is generally weak and multi-photon events occur with extremely low probability. Consequently, it is extremely challenging to engineer quantum nonlinear devices that operate at the single photon level and the majority of quantum technologies have to rely on single photon interactions. Here, we demonstrate experimentally and theoretically that exploiting coherent absorption of $N=2$ N00N states makes it possible to enhance the number of two-photon states that are absorbed. An absorbing metasurface placed inside a Sagnac-style interferometer into which we inject an $N = 2$ N00N state, exhibits two-photon absorption with 40.5$\%$ efficiency, close to the theoretical maximum. This high probability of simultaneous absorption of two photons holds the promise for applications in fields that require multi-photon upconversion but are hindered by high peak intensities.}
\end{abstract}

\maketitle

\section{Introduction}

\begin{figure*}
\centering
\includegraphics[width = \textwidth]{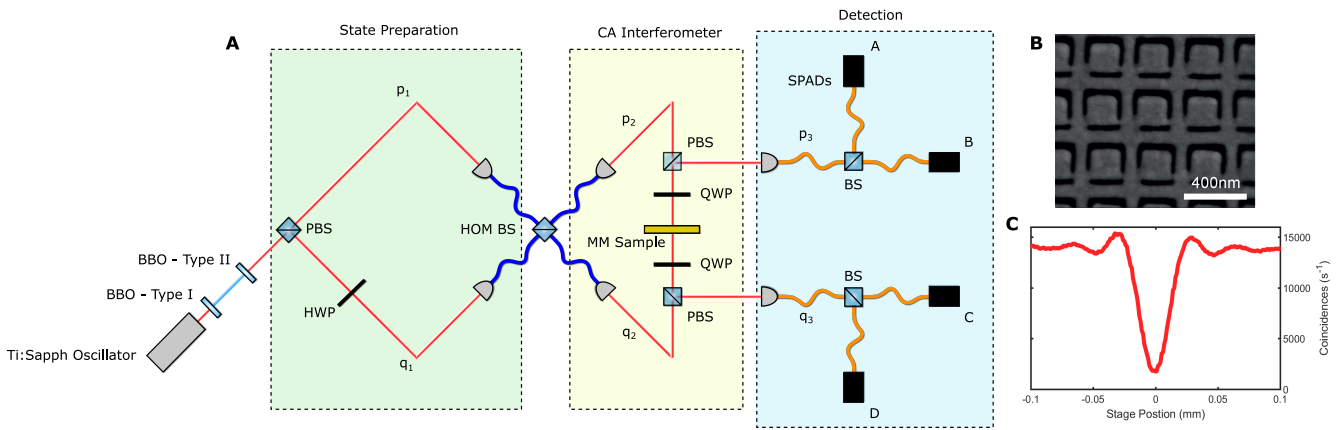}
\caption{A) Schematic of the experiment split into three sections. The ``State Preparation" section creates the N = 2 N$00$N states using  HOM interference. The second section (``Coherent Absorption Interferometer") is a modified Sagnac interferometer using circular polarisation at the Meta Material Sample (MM Sample). Light is coupled out of the ``CA Interferometer" by rotating back to linear polarisation and using Polarising Beamsplitters (PBS) to achieve a high extraction efficiency. Finally, the photons are counted in the ``Detection" section which consists of fiber-coupled 50:50 beamsplitters (BS) and four Single Photon Avalanche Diodes (SPADs). B) SEM image of the meta-material made up of an array of split ring resonators milled into a 50 nm thick free-standing gold film by a Focused Ion beam. C) The HOM dip as measured at the output ports of the HOM BS by translating one of the fiber-couplers at one of the input ports along the beam propagation direction.}
\label{exp_layout_fig}
\end{figure*}

The efficiency with which multiple photons can be instantaneously absorbed by a medium typically has a nonlinear dependence on the amplitude of the optical fields and is therefore negligible at the few-photon level. This is in general true for all forms of photon-photon interaction in the quantum (few photon) domain when compared, for example, to high power, classical nonlinear optics where a range of multi-photon effects are readily available (multi-photon absorption, saturable absorption, frequency conversion etc.) and are therefore currently used in a range of devices.  
So, although schemes to produce deterministic multi-photon states have been proposed \cite{Morandotti}, in general, the lack of access to quantum multi-photon processes severely inhibits the use of optics for a large number of applications surrounding quantum technologies. For instance, many quantum logic gates require the interaction between two qubits which could be moderated by the absorption of two photons \cite{Kok2007}. Multi-photon absorption at the few photon level could also lead to photon counting methods with a much higher level of certainty than current techniques \cite{Lusardi2017,Divochiy2008,Dauler2009} that also require the identification of events where exactly a set number of photons are absorbed. In addition, quantum states of light have been proposed as a method of surpassing the diffraction limit for photolithography but this is highly reliant on the simultaneous absorption of a specific number of photons \cite{Boto2000,Dowling2008}. A mechanism to enhance quantum multiphoton processes would therefore be highly desirable.\\
Here we demonstrate a route to highly efficient quantum multiphoton processes relying on Coherent Perfect Absorption (CPA), a process that exploits the coherent properties of light to achieve complete absorption \cite{Chong2010,Longhi2010,Baranov2017}. Studies have recently shown how CPA can be achieved with ultra-thin materials \emph{i.e.} with a thickness much less than the wavelength of the light. CPA in such 2D media  has been demonstrated not only with continuous wave sources \cite{Zhang2012a} but also with pulsed sources on the femtosecond timescale \cite{Rao2014} and has been extended to investigations into the coherent control of nonlinear effects such as four-wave mixing \cite{Rao2015}. 
Although in what follows we will discuss the quantum phenomenon of two-photon absorption in a thin absorber, it is worth noting that the absorber with no intrinsic nonlinearity that interacts with two incident waves and two outgoing waves is a four port device that can provide nonlinear input-output signal dependencies even in the classical regime of wave interactions. Such nonlinear response results from redistribution of energy among ports without introducing signal distortion and may be used to small-signal amplifier, summator and invertor functions for optical signals \cite{Fang2015}. For instance,  at a constant intensity and zero phase difference between input and pump waves, the output light intensity will be a nonlinear function of input light intensity.\\
Further studies have started to reveal how CPA interacts with quantum states of light and experiments using single photons have shown that deterministic complete absorption or complete transmission of the photons can be achieved \cite{Roger2015, Huang2014}. However, very few investigations have addressed the coherent absorption of quantum multi-photon states, with just a handful of examples available in the literature \cite{Jeffers2000, Barnett1998, Roger2016, Vest2017}. \\
It was shown in a recent proof-of-principle experiment that two-photon states may exhibit the effects of CPA where a weak modulation in the coincidence count rate was observed, with graphene acting as the thin absorbing medium \cite{Roger2016}. Two main factors stand out in this work: the weak modulation that is only of order of a few percent and the lack of a protocol that would enable one to quantitatively estimate the efficiency of the two-photon absorption process. The latter issue is due to the inability for any device to directly measure zero-photon states.
Overall, the weak modulation in the coincidence counts indicates that the efficiency in this experiment was far from the theoretical limit of 50\%, predicted by theoretical studies \cite{Jeffers2000, Barnett1998} and thus fails to provide evidence that high absorption probabilities are achievable. Naturally, this will continue to limit the possibilities where coherent absorption of quantum states can be employed, for example, for photonic logic gates \cite{Fang2015}. 
Provided that efficiencies on the scale of this 50\% limit can be demonstrated, CPA can provide a pathway to two-photon absorption schemes which are not accessible with classical states of light.\\
In this work we demonstrate for the first time, that absorption probabilities on the scale of this fundamental 50\% limit are possible. To achieve this, we use a bespoke metamaterial with absorption, reflection, and transmission coefficients close to the optimum values. We also introduce a measurement protocol which quantifies the amount of two-photon absorption from a simple measurement of only the rate of coincidence events between pairs of detectors with no free fitting parameters. By using this model we demonstrate two-photon coherent absorption with 40\% efficiency, the highest ever reported. In fact, the results presented here prove that the theoretical limit on coherent absorption of multiphoton states can be approached and pave the way for using two-photon coherent absorption in various applications ranging from quantum gates to probing quantum coherence in biological molecules.\\
 
\section{Theory}

Our results rely on the use of an important class of multi-photon quantum states, known as N00N states, which have the form $\ket{\Psi_{\text{N00N}}} = {1}/{\sqrt{2}} \left( \ket{\text{N}_{\text{a}}, 0_{\text{b}}} + e^{- i \text{N} \theta} \ket{0_{\text{a}}, \text{N}_{\text{b}}} \right)$ (where N is the number of photons, a \& b represent two possible modes that the photons can occupy and $\theta$ is an arbitrary phase shift). It can be shown that, for maximum efficiency CPA in an infinitely thin film with a linear absorption $\alpha = 0.5$, the reflection and transmission coefficients must be equal in magnitude with a relative phase between them that is equal to $0$ \cite{Baranov2017, Fang2015, Dutta-Gupta2012, Zhang2012a, Thongrattanasiri2012}. A good approximation to this scenario can be engineered using meta-materials with thickness, $d\ll\lambda$ where $\lambda$ is the wavelength of the photons. 
\vspace{0.1cm}
\hspace{-0.2cm}
\begin{table}
\centering
\begin{tabular}{| p{2.0cm} p{1cm} p{4.5cm} |} 
\hline
\vspace{0.0cm}$r$ \& $t$ Phase &\vspace{0cm} Input &\vspace{0cm} Output\\
\hline
\hline
\vspace{0.0cm}
0 & \vspace{0.1cm} $\ket{2_+}$ & \vspace*{0.0cm} $ \sfrac{1}{2} ( \ket{0_a,0_b} \bra{0_a,0_b} + \ket{\psi_2} \bra{\psi_2} ) $  \vspace{0.0cm}\\
0 & $\ket{2_-}$ & $\mp \ket{1_\pm}$ \vspace{0.1cm}\\
\hline
\end{tabular}
\caption{Output states from a beamsplitter with 50\% loss for the two opposite input N00N phases, $\theta=0$ (indicated as the $\ket{2_+}$ input state)  and $\theta=\pi$ (indicated as the $\ket{2_-}$ input state).   $\ket{\psi_2} = \sfrac{1}{2}\left(\ket{1_{\text{a}},1_{\text{a}}} + \ket{2_{-}} \right)$.
}
\label{state_table}
\end{table}
The output states from such a metamaterial beamsplitter with an N = 2 N00N input state, are summarised in  Table \ref{state_table} \cite{Jeffers2000}: by varying the N00N phase $\theta$, the output states change from a single photon state to a mixture of two-photon states and zero-photon states (where both photons are absorbed). The latter implies that, for the correct phase of the interferometer ($\theta = 0$, $2\pi$, $4\pi$ \ldots), one can expect 50\% simultaneous absorption of both photons. Note that the time averaged number of photons measured per input pair will always be 1, therefore more sophisticated techniques need to be employed in order to fully demonstrate this effect. In other words, photon counters cannot  directly provide the amplitude of the zero-photon state. We show that this can however be achieved by measuring the amplitudes of the output two-photon states together with the system losses, and then inferring the zero-photon states from these measurements. \\
We build a theoretical model which accounts for the three main processes in the experiment: (i) The generation of two correlated single photons via spontaneous parametric down conversion, (ii) the formation of an N=2 N00N state by combining the generated photons onto a lossless beamsplitter and (iii) the interaction of the N=2 N00N state with the lossy metamaterial beamsplitter and the final measurement of the resultant coincident photon counts. In more detail, the input density matrix $\rho_0$ evolves according to
\begin{align}
\rho & = \sum_{i,j,k=0}^{5} F_{k}\mathcal{U}_{2}E_{j}^{p_2,q_2} \mathcal{U}_{1} E_{i}^{p_1,q_1} \rho_{0} E_{i}^{p_1,q_1\dagger} \mathcal{U}_{1}^{\dagger} E_{j}^{p_2,q_2\dagger} \mathcal{U}_{2}^{\dagger} F^{\dagger}_{k}
\label{final_rho}
\end{align}
where the indices $i,j,k$ indicate the sum over the dimension of the noise Hilbert space (used to describe the trivial and coherent losses, see supplemental information). The input SPDC state is $\rho_0 = |11\rangle\langle11|$, where we use the ordered basis to describe states with up to two photons $|00\rangle$, $|01\rangle$, $|10\rangle$, $|11\rangle$, $|20\rangle$, $|02\rangle $. $E_{i}, E_{j}$  are the operators accounting for system losses (absorption, reflection from optical components, scattering from imperfections etc.) between the SPDC crystal and HOM BS, and then between the HOM BS and the metamaterial film, respectively. The unitary operator $\mathcal{U}_{1}$ describes the lossless HOM BS, while $\mathcal{U}_{2}$ is the unitary propagation operator describing the phase shift in the interferometer. Finally, $F_{k}$ are the operators describing the lossy metamaterial BS. The parameters in the loss operators $E_{i}$ are derived from the measured transmission of each individual interferometer arm, as described in more detail in the Supplementary Materials. It is then possible to calculate the coefficient of the $|00\rangle$ basis vector from measurements at the metamaterial output, as described below.\\
%
\section{Experimental Setup}

\begin{figure}
\includegraphics[width = 0.5\textwidth]{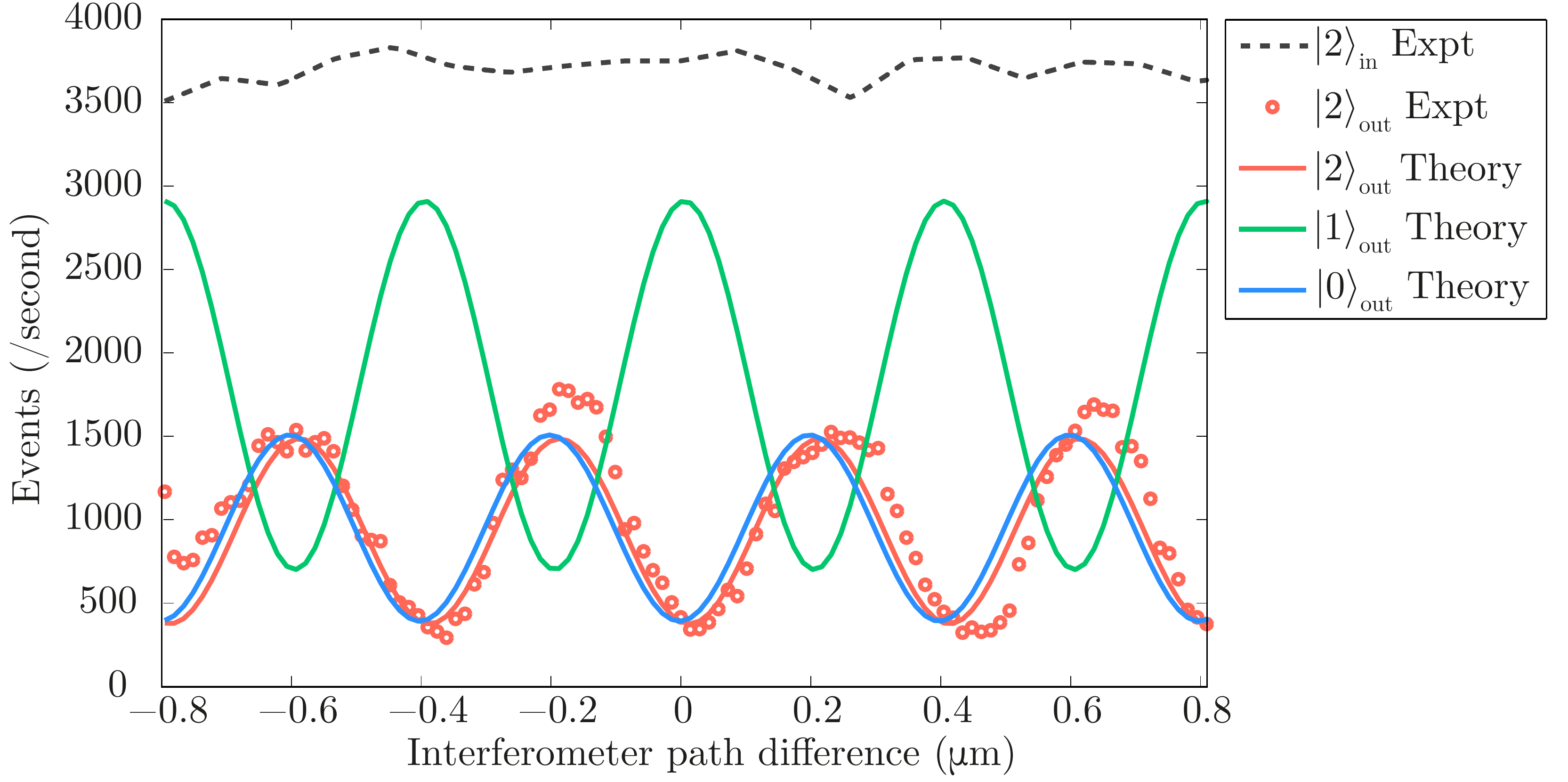}
\caption{Amplitudes  of zero, one, and two-photon states as a function of the interferometer phase. The plots are based on the experimentally measured values of all coincidence-count combinations, incorporating the losses at each stage of the setup. The number of input N00N states is estimated from the total number of coincident counts measured without the absorbing sample and the (separately mmeasured) HOM visibility (Fig.~\ref{exp_layout_fig}c). We note the good agreement between the measured $|2\rangle_\textrm{out}$ and the theoretical estimate of the same quantity, with no free parameters, indicating that the model indeed describes the setting well. }
\label{losses_fig}
\end{figure}

The experimental layout consists of two cascaded interferometers shown in Fig.~\ref{exp_layout_fig}a and described in more detail in the Supplementary Materials.
The first interferometer, based on a lossless BS, generates the N00N states and these are sent to the second ``Coherent Absorption'' (CA) interferometer. This consists of two counter-propagating arms which meet at the free-standing metamaterial sample (see Fig. \ref{exp_layout_fig}b and supplementary information for the details of the metamaterial film) in the same fashion as a Sagnac interferometer, which is modified to allow for complete coupling out of the interferometer via polarising beamsplitters (PBS) and quarter waveplates (QWP). The use of circular polarisation has minimal effect in terms of the measured two-photon state absorption. Both of the output ports are then coupled to multimode fibers and sent via fiber-coupled beamsplitters to four SPAD detectors, which we label A, B, C \& D. Previous experiments using plasmonic interactions (such as those utilised by our metamaterial) have demonstrated that they can be used for quantum interference and entanglement experiments without significant decoherence of the state \cite{Fakonas2014,Fakonas2015,Vest2017}.\\ 

\section{Results}
The total number of two-photon states is determined by measuring the total number of two-fold coincident counts between all possible combinations of detectors A, B, C \& D. In order to provide an accurate estimate, there are two considerations which must be taken into account. The first is due to the probabilistic nature of the beamsplitters used for detection which will only separate 50\% of the incident two-photon states. The total number of two-photon states is therefore,
\begin{equation}
\text{C}_{\text{Total}} = 2 \left[ \text{C}_{\text{AB}} + \text{C}_{\text{CD}} \right] + [\text{C}_{\text{AC}}+\text{C}_{\text{AD}}+\text{C}_{\text{BC}}+\text{C}_{\text{BD}}],
\label{coin_eq}
\end{equation}
where $\text{C}_{\text{ij}}$ represents the number of coincident counts between detectors A, B, C \& D as indicated in the subscripts i \& j. The second consideration is the loss throughout the system due to fiber coupling efficiency and non-unity transmission of the components. This is accounted for by the $E_i$ matrices in Eq.~\ref{final_rho}. In order to recover the output state from measurements after the metamaterial beamsplitter, we model the system by Eq.~\ref{final_rho}. The experimentally measured coincidence rates are then compared to those predicted by our theoretical model. We do this by renormalising the rates according to Eq.~\ref{coin_eq} and by accounting for the measured detector and fibre coupling efficiencies ($\eta_{detector} = 0.62$ and $\eta_{coupling} = 0.7$ respectively). 
Fig.~\ref{losses_fig} shows the total coincidence  counts $\text{C}_{\text{Total}}$ corresponding to two-photon states (circles)  as the phase of the interferometer is varied, and compares this to the corresponding theoretical estimate (red curve). The good agreement with no free parameters indicates the high fidelity of the model.
\begin{figure}
\centering
\includegraphics[width = 0.4\textwidth]{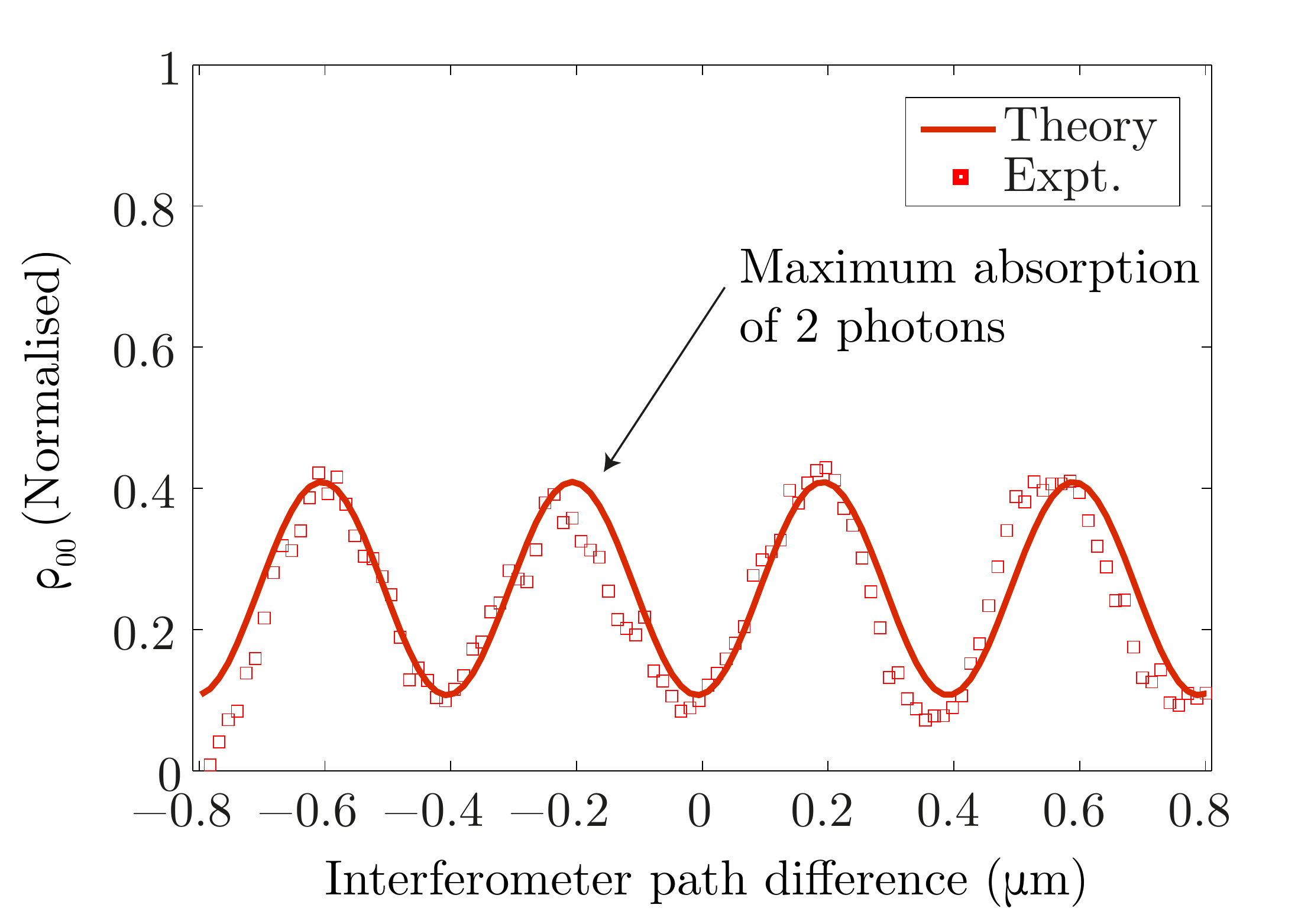}
\caption{Variation of the 2 photon absorption coefficient ($\rho_{00}$) as a function of the interferometer path difference, measured at the metamaterial sample. The squares represent the experimentally measured component, which has been interpreted using the theoretical model including losses. The solid line represents the theoretically calculated $\rho_{00}$ component with no free parameters.} 
\label{2photon}
\end{figure}
The amount of two-photon absorption is  then quantified directly from the output density matrix Eq.~\ref{final_rho} projected onto the basis vector $|00\rangle$. A similar operation can be performed for the one-photon states. These results are plotted in Fig.~\ref{losses_fig} along with the total two-photon input rate $|2\rangle_\textrm{in}$ measured at the HOM BS output (dashed line). The ratio between $|0\rangle_\textrm{out}$ and $|2\rangle_\textrm{in}$ provides a direct estimate of the total absorption of the two-photon state and is found to reach a maximum of 40.5\% (periodic with the interferometer phase). This approaches the theoretical maximum of 50\% (see supplemental information for more details). This result is shown differently in Fig.~\ref{2photon}  where we plot the normalised zero-photon state coefficient $|0\rangle_\textrm{out}=(|2\rangle_\textrm{in}-|1\rangle_\textrm{out} -|2\rangle_\textrm{out})/|2\rangle_\textrm{in}$ together with the theoretical prediction (where all quantities in the formula are evaluated from the theoretical model).  \\
Finally, in Fig.~\ref{noon_enh_fig} we show the expected enhancement of the absorption for a range of N00N states (N=1 to N=7) for the parameters of the metamaterial used in the experiments and for the ideal case of $\alpha = 0.5$. In the case of the simple N independent-photon absorption the probability of absorbing all N photons scales as the absorption, $\gamma$, of one photon to the Nth power, i.e. $\text{p}(\ket{1,0}^\text{N}) = \gamma^\text{N}$. On the other hand, for N00N states $\text{p}(\text{N}00\text{N}) = \gamma^\text{N} + \delta^\text{N}$ where $\delta = \text{rt}^* + \text{r}^*\text{t}$. There is therefore always an improvement of a minimum factor $2\times$ using N00N states. The results of this work for N=2 can therefore be generalised to higher order N00N states without loss of generality.

\begin{figure}
\centering
\includegraphics[width = 0.45\textwidth]{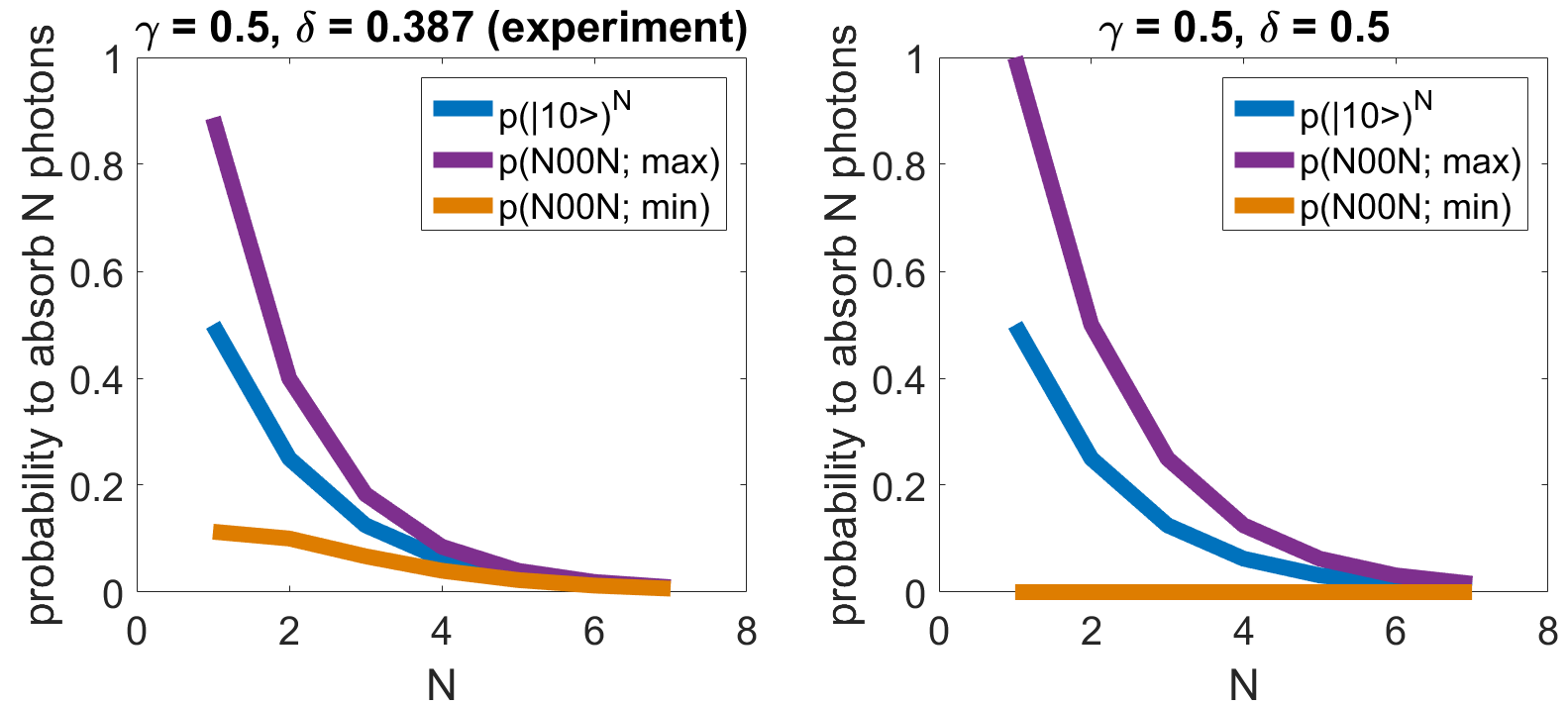}
\caption{Enhancement and predicted controllability of the absorption probability of exactly N photons from a N00N state (max - red, min - yellow) compared to N independent photons (blue). A) Parameters for the metasurface used in the experiment. B) Parameters for maximum enhancement ($\gamma = 0.5$).}
\label{noon_enh_fig}
\end{figure}

\section{Conclusions}

In summary, we have provided experimental evidence of efficient coherent two-photon interaction, in the form of two-photon state absorption in a metamaterial. The two-photon absorption rate is quantified through a specifically designed measurement protocol and was recorded to be close to the theoretical maximum of 50\%. We note that we only need to measure coincidence rates between the four detectors and characterisation of the individual optical components (i.e. losses and transmission), to yield the full density matrix - thereby providing the coefficients of all possible output states composed of zero, one and two photons. With no free parameters, the model is able to reproduce the total coincidence rates measured in the experiment. This allows us to retrieve the zero-photon amplitudes, even without a method to characterise the one-photon amplitudes directly.\\
The mechanism of absorption observed here is via conversion to localised plasmons. There have also been a number of recent quantum optics studies that couple photons to travelling plasmons in order to study their coherence properties. N00N state coherent absorption would be a route to efficient generation of travelling plasmon Fock-states with coherent control over the plasmon number (one or two) by changing the interferometer phase. \\
  Another opportunity could rely on the temporal entanglement that is shared by the signal and idler photons produced by SPDC, which can give rise to an increase in the efficiency with which both photons can be absorbed by systems requiring pumping on an ultrafast timescale yet have very narrowband transitions \cite{Schlawin2013}. The efficiency of such effects could potentially be enhanced to yield an absorption cross-section close to unity by combining them with the coherent mechanism demonstrated here.\\

\section*{Funding Information}
D.F. acknowledges support from the European Research Council under the European Union’s Seventh Framework Programme (FP/2007-2013)/ERC, Grant No. GA 306559, the Engineering and Physical Sciences Research Council (EPSRC, UK, Grants No. EP/M006514/1 and No. EP/M01326X/1) and the Leverhulme Trust.

\section*{Acknowledgments}

\bigskip \noindent See \href{link}{Supplement 1} for supporting content.

\bibliographystyle{unsrt}

\clearpage

\onecolumngrid
\Large
Supplemental Material: Coherent metamaterial absorption of two-photon states with 40\% efficiency.
\vspace{1cm}
\twocolumngrid
\normalsize

\section{Experimental layout and metamaterial sample} The experimental layout consists of two cascaded interferometers, as shown in Fig. 1a (main document). The first interferometer, based on a lossless BS, is used to generate the N00N states. The N00N states are prepared using wavelength-degenerate (808 nm) spontaneous parametric down conversion (SPDC) within a type II BBO (beta-barium borate) crystal pumped by a modelocked Ti:Sapph oscillator (Coherent Inc. Chameleon Ultra II), that has been frequency doubled via type I BBO to provide 404 nm pulses of light with 130 fs duration at 80 MHz repetition rate. The signal and idler photons are then split using a polarising beam splitter (PBS). The polarisation of the idler photon is then rotated to match that of the signal. The photons are individually coupled into polarisation maintaining fiber, where they are guided to a 50:50 cube beamsplitter and undergo two photon interference \cite{Hong1987}. The HOM effect ensures that the photon pair leave via the same output port of the beamsplitter in a superposition state, i.e. in an N $=$ 2 N00N state \cite{Dowling2008}. The HOM dip visibility is measured to be $78 \%$ (see Fig.  1c). 
The second ``Coherent Absorption'' (CA) interferometer  consists of two counter-propagating arms which meet at the free-standing metamaterial sample,  in the same fashion as a Sagnac interferometer but modified to allow for complete coupling out of the interferometer via PBSs and quarter waveplates. A linearly polarised photon passes through the PBS and is transformed into a circularly polarised photon. Then, the photon passes through the metamaterial, is re-converted back to a linearly polarised photon (with polarization orthogonal to that of the original state) and is therefore coupled out from the interferometer with close to 100 \% efficiency by the second PBS. This setting allows complete extraction of photons from the interferometer, with minimal loss of information.\\
Both of the output ports are then coupled into multimode fibers and sent via fiber-coupled beamsplitters to four SPAD detectors, which we label A, B, C \& D. The four SPAD detectors allow for counting of up to two photons in each output mode. Counting of the two-photon states is then achieved by time-tagging all detected photon events. \\
The `lossy beamsplitter' used here is a freestanding gold film with a split-ring resonator array milled via a focused ion beam creating sub-lambda resonators (see Fig.  1b). We measure the reflection and transmission coefficients of the sample (free standing, without the HOM setup) using circularly polarised light to be $t = 0.66$ and $r = 0.2921 - i0.217$.\\

\section{Experimental Component characterisation} The matrices $E_i$ in Eq. 1 (main text), describing the evolution of the photons under trivial losses, require knowledge of the losses through each arm of the HOM and CA interferometers. These were determined experimentally by measuring the ratio between input and output single photon counts for each arm. These values are indicated in Tab.~\ref{table}, where $q_i$ and $p_i$ refer to the interferometer arms as indicated in Fig. 1.
\begin{table}[ht!]
\begin{center}
\begin{tabular}{| c c c |}
    \hline
    {\bf Component} & {\bf Parameter} & {\bf Transmission} \\ 
    \hline
    \hline
   HOM BS & $p_1$ & 0.39 \\ 
     & $q_1$ & 0.19 \\ 
    Intermediate loss & $p_2$ & 0.621 \\ 
     & $q_2$ & 0.626 \\ 
     Output loss & $p_3$ & 0.7 \\ 
     & $q_3$ & 0.6 \\ 
    \hline
 \end{tabular}
 \end{center}
 \caption{\label{table} Table showing measured transmission of interferometer arms.}
 \end{table}
 
 \section{Theoretical model}

\subsection{Quantum Evolution with Loss}

The system described in the main text consists of a two photon input state propagating in an optical setup. The setup includes linear optical components (beamsplitters and phase shifters), as well as a lossy beamsplitter (LBS). Furthermore, the setup includes regular incoherent loss, as opposed to the quantum coherent loss at the LBS. Due to the photon losses, the system cannot be described by a unitary evolution operator acting on the state. Instead, we will find the Kraus operators corresponding to the propagation in the various elements \cite{nielsen2002quantum}. Kraus operators $\left\{E_i\right\}_{i=1}^{n}$ describing a trace preserving quantum operation satisfy

\begin{equation}
\sum_{i}E_{i}^{\dagger}E_{i} = I,
\end{equation}

where the evolution of the state, described by the densty matrix $\rho$, is given by

\begin{equation}
\rho \mapsto \sum_{i}E_{i}\rho E_{i}^{\dagger}.
\end{equation}

To obtain the operators, we will utilize our knowledge on the single photon propagation in the lossy and lossless elements. In the lossy elements, the evolution involves extra noise modes. The evolution of all the modes $\mathcal{V}$, including the noise modes, is unitary $\mathcal{V}\mathcal{V^{\dagger}} = I$. However, since we do not have control over the noise modes, we will trace over the noise degrees of freedom.\\

Suppose the initial state of the system is $\rho$, and the initial state of the noise modes is $\vert n_0 \rangle$. The evolution of the original system is given by

\begin{align}
\rho' & = \text{Tr}_{\text{noise}} [ \mathcal{V} ( \rho \otimes \vert n_0 \rangle \langle n_0 \vert ) \mathcal{V^{\dagger}}] \\
        & = \sum_{i} \langle e_i \vert \mathcal{V} ( \rho \otimes \vert n_0 \rangle \langle n_0 \vert ) \mathcal{V^{\dagger}} \vert e_i \rangle \\
        & = \sum_{i} E_{i} \rho E_{i}^{\dagger}.
\end{align}

Here, $\{ \vert e_i \rangle \}$ is an orthonormal basis of the noise Hilbert space.

\subsection{Propagation in the Interferometer}

In this section we show how to include loss in the quantum optics description of the N = 2 N00N state CPA experiment. The loss could be due to coupling into fiber or absorption within an optical element. This is in contrast to the loss inherent in the `lossy beamsplitter' approach, which is included within the evolution through the metamaterial sample and explained in the following section.\\

Suppose we have two modes $a_1^{\dagger}$ and $a_2^{\dagger}$, corresponding to the two arms of the interferometer (see Fig. \ref{fig:scheme}). Each mode will experience loss in the following manner.

\begin{align}
a_1^{\dagger} &\rightarrow \sqrt{1-p}a_1^{\dagger} + \sqrt{p}c^{\dagger},\\
a_2^{\dagger} &\rightarrow \sqrt{1-q}a_2^{\dagger} + \sqrt{q}d^{\dagger},
\end{align}

where $p$, $q$ are the probabilities with which a photon is lost in channels 1 and 2 respectively. The noise operators $c^{\dagger}$, $d^{\dagger}$ describe photon creation in the loss modes, which are independent of each other and of the rest of the system. These operators are identical to the photons in that they satisfy the same commutation relations. That is
\begin{equation}
[c, c^{\dagger}] = [d, d^{\dagger}] = 1,
\end{equation}

\noindent all other commutation relations are equal to zero.\\

We start with $N=2$ photons in the system (modes $1,2$) and no photons in the noise modes $\vert n_0 \rangle = \vert 00\rangle_{c,d}$. The total number of photons ($N=2$) is conserved, however, the partition between the original system and noise modes can change. Therefore, we need bases consisting of $6$ basis vectors for each of the Hilbert spaces, describing up to $2$ photons in each. As in the main text, we assume the ordered orthonormal basis ($|00\rangle$, $|10\rangle$, $|01\rangle$, $|11\rangle$, $|20\rangle$, $|02\rangle$). Under the unitary evolution $\mathcal{V}_1$, an initial state with no photons in the noise modes and one of the original system basis vectors, propagates in the following manner:

\begin{align*}
|00\rangle |00\rangle &\rightarrow |00\rangle |00\rangle ,\\
|10\rangle |00\rangle &\rightarrow \sqrt{1-p}|10\rangle|00\rangle + \sqrt{p}|00\rangle|10\rangle ,\\
|01\rangle |00\rangle &\rightarrow \sqrt{1-q}|01\rangle|00\rangle + \sqrt{p}|00\rangle|01\rangle ,\\
|11\rangle |00\rangle &\rightarrow \sqrt{(1-p)(1-q)}|11\rangle|00\rangle + \sqrt{q(1-p)}|10\rangle|01\rangle \\
                                & +\sqrt{p(1-q)}|01\rangle|10\rangle + \sqrt{pq}|00\rangle|11\rangle ,\\
|20\rangle |00\rangle &\rightarrow (1-p)|20\rangle |00\rangle + \sqrt{2p(1-p)}|10\rangle|10\rangle \\ &+ p|00\rangle|20\rangle ,\\
|02\rangle |00\rangle &\rightarrow (1-q)|02\rangle |00\rangle + \sqrt{2q(1-q)}|01\rangle|01\rangle\\& + q|00\rangle|02\rangle.
\end{align*}

Tracing over the noise degrees of freedom yields the following quantum operation:

\begin{equation}
E_0 = 
\begin{pmatrix}
  1 & 0 & 0 & 0 & 0 & 0 \\
  0 & \sqrt{1-p} & 0 & 0 & 0 & 0 \\
  0 & 0 & \sqrt{1-q} & 0 & 0 & 0 \\
  0 & 0 & 0 & \sqrt{(1-p)(1-q)} & 0 & 0 \\
  0 & 0 & 0 & 0 & 1-p & 0 \\
  0 & 0 & 0 & 0 & 0 & 1-q \\
 \end{pmatrix}
 \end{equation}
 
 \begin{equation}
E_1 = 
\begin{pmatrix}
  0 & \sqrt{p} & 0 & 0 & 0 & 0 \\
  0 & 0 & 0 & 0 &\sqrt{2p(1-p)} & 0 \\
  0 & 0 & 0 & \sqrt{p(1-q)} & 0 & 0 \\
  0 & 0 & 0 & 0 & 0 & 0 \\
  0 & 0 & 0 & 0 & 0 & 0 \\
  0 & 0 & 0 & 0 & 0 & 0\\
 \end{pmatrix}
 \end{equation}
 
  \begin{equation}
E_2 = 
\begin{pmatrix}
  0 & 0 & \sqrt{q} & 0 & 0 & 0 \\
  0 & 0 & 0 & \sqrt{q(1-p)} & 0 & 0 \\
  0 & 0 & 0 & 0 & 0 & \sqrt{2q(1-q)} \\
  0 & 0 & 0 & 0 & 0 & 0 \\
  0 & 0 & 0 & 0 & 0 & 0 \\
  0 & 0 & 0 & 0 & 0 & 0\\
 \end{pmatrix}
 \end{equation}
 
   \begin{equation}
E_3 = 
\begin{pmatrix}
  0 & 0 & 0 & \sqrt{pq} & 0 & 0 \\
  0 & 0 & 0 & 0 & 0 & 0 \\
  0 & 0 & 0 & 0 & 0 & 0 \\
  0 & 0 & 0 & 0 & 0 & 0 \\
  0 & 0 & 0 & 0 & 0 & 0 \\
  0 & 0 & 0 & 0 & 0 & 0\\
 \end{pmatrix}
 \end{equation}
    \begin{equation}
E_4 = 
\begin{pmatrix}
  0 & 0 & 0 & 0 & p & 0 \\
  0 & 0 & 0 & 0 & 0 & 0 \\
  0 & 0 & 0 & 0 & 0 & 0 \\
  0 & 0 & 0 & 0 & 0 & 0 \\
  0 & 0 & 0 & 0 & 0 & 0 \\
  0 & 0 & 0 & 0 & 0 & 0\\
 \end{pmatrix}
 \end{equation}
    \begin{equation}
E_5= 
\begin{pmatrix}
  0 & 0 & 0 & 0 & 0 & q \\
  0 & 0 & 0 & 0 & 0 & 0 \\
  0 & 0 & 0 & 0 & 0 & 0 \\
  0 & 0 & 0 & 0 & 0 & 0 \\
  0 & 0 & 0 & 0 & 0 & 0 \\
  0 & 0 & 0 & 0 & 0 & 0\\
 \end{pmatrix}
 \end{equation}
 
where indeed $\sum_{i=0}^{5}E_i^{\dagger}E_i = I$. \\

We may therefore model the first part of our experiment i.e. the interaction with the lossless beamsplitter and loss at $p_1$, $q_1$ (coupling losses into the fibers, see Fig. \ref{fig:scheme}), by evolving the density matrix according to

\begin{equation}
\rho  \rightarrow \sum_i\mathcal{U}_{1}E_i\rho E_i^{\dagger}\mathcal{U}^{\dagger}_{1},
\end{equation}

where $\mathcal{U}_{1}$ is the unitary evolution of the lossless beamsplitter.\\

Next, the photons propagate through the interferometer with losses $p_2, q_2$, and channel $2$ experiences a phase shift of $\varphi$, mimicking the relative shift of the Sagnac interferometer path length. Thus, the state right before the metamaterial (the LBS), is given by

\begin{align}
\rho_1 & = \sum_{i,j=0}^{5} \mathcal{U}_{2}E_{j}^{p_2,q_2} \mathcal{U}_{1} E_{i}^{p_1,q_1} \rho E_{i}^{p_1,q_1\dagger} \mathcal{U}_{1}^{\dagger} E_{j}^{p_2,q_2\dagger} \mathcal{U}_{2}^{\dagger},
\label{rho1}
\end{align}

where $\mathcal{U}_{2}$ is the phase shift $\varphi$, and $E_{i}^{p,q}$ is the $i$th operator in the quantum operation describing independent losses $p,q$.

\subsection{Quantum Operation of a Lossy Beamsplitter}

Here we describe the action of an arbitrary `lossy beamsplitter' (LBS). The evolution of a mode upon interaction with the LBS is given by \cite{Barnett1998}

\begin{eqnarray}
a_1^{\dagger} \rightarrow ta_1^{\dagger} + ra_2^{\dagger} + c^{\dagger},\\
a_2^{\dagger} \rightarrow ra_1^{\dagger} + ta_2^{\dagger} + d^{\dagger},
\end{eqnarray}

where $r,t$ are the reflection and transmission coefficients of the LBS. In this case, the noise operators $c^{\dagger}$, $d^{\dagger}$ no longer describe independent particles. Instead, they obey the commutation relations

\begin{eqnarray}
[c, c^{\dagger}] = [d, d^{\dagger}] = 1 - |t|^2 - |r|^2, \\
\left[c, d^{\dagger}\right] = [d, c^{\dagger}] = -tr^* - rt^*.
\end{eqnarray}

The unitary evolution $\mathcal{V}_{2}$ of the total system (original modes and noise modes), assuming the noise modes are initially unpopulated, is given by

\begin{align}
|00\rangle |00\rangle &\rightarrow |00\rangle |00\rangle ,\\
|10\rangle |00\rangle &\rightarrow t|10\rangle|00\rangle + r|01\rangle|00\rangle + |00\rangle c^{\dagger}|00\rangle ,\\
|01\rangle |00\rangle &\rightarrow r|10\rangle|00\rangle + t|01\rangle|00\rangle + |00\rangle d^{\dagger}|00\rangle ,\\
|11\rangle |00\rangle &\rightarrow (ta_1^{\dagger}+ra_2^{\dagger} + c^{\dagger})(ra_1^{\dagger}+ta_2^{\dagger} + d^{\dagger})|00\rangle|00\rangle ,\\
|20\rangle |00\rangle &\rightarrow \frac{1}{\sqrt{2}}(ta_1^{\dagger}+ra_2^{\dagger} + c^{\dagger})^2|00\rangle|00\rangle ,\\
|02\rangle |00\rangle &\rightarrow \frac{1}{\sqrt{2}}(ra_1^{\dagger}+ta_2^{\dagger} + d^{\dagger})^2|00\rangle|00\rangle .
\end{align}

To obtain the quantum operation, we will find an orthonormal basis of the noise Hilbert space. We define the operators

\begin{align}
C^{\dagger} & = \frac{1}{\sqrt{2(1 - \vert t + r \vert^{2})}}(c^{\dagger} + d^{\dagger}), \label{noise_operators_transformation1}\\
D^{\dagger} & = \frac{1}{\sqrt{2(1 - \vert t - r \vert^{2})}}(c^{\dagger} - d^{\dagger}).
\label{noise_operators_transformation2}
\end{align}

They satisfy the canonical commutation relations:

\begin{align}
[C,C^{\dagger}] &= [D,D^{\dagger}] = 1, \\
[C,D^{\dagger}] &= [D,C^{\dagger}] = 0. 
\end{align}

Therefore, the following basis of the noise Hilbert space is orthonormal:

\begin{align}
\vert 00 \rangle ,\\
\vert 10 \rangle & = C^{\dagger}\vert 00 \rangle ,\\
\vert 01 \rangle & = D^{\dagger}\vert 00 \rangle ,\\
\vert 11 \rangle & = C^{\dagger}D^{\dagger}\vert 00 \rangle ,\\
\vert 20 \rangle & = \frac{1}{\sqrt{2}}C^{\dagger 2}\vert 00 \rangle ,\\
\vert 02 \rangle & = \frac{1}{\sqrt{2}}D^{\dagger 2}\vert 00 \rangle. 
\end{align}

\begin{widetext}
Using this basis, denoted by $\{|e_i \rangle\}_{i=0}^{5}$, and the transformation of the noise operators (Eq. \ref{noise_operators_transformation1}, \ref{noise_operators_transformation2}), we can trace over the noise modes to obtain the quantum operation of the LBS $F_{i} = \langle e_i | \mathcal{V}_{2} | 00\rangle$, with $\sum_{i=0}^{5} F_i^{\dagger}F_i = I$, and LBS parameters $r,t$:

\begin{equation}
F_0 = 
\begin{pmatrix}
  1 & 0 & 0 & 0 & 0 & 0 \\
  0 & t & r & 0 & 0 & 0 \\
  0 & r & t & 0 & 0 & 0 \\
  0 & 0 & 0 & t^2 + r^2 & \sqrt{2}tr & \sqrt{2}tr \\
  0 & 0 & 0 & \sqrt{2}tr & t^2 & r^2 \\
  0 & 0 & 0 & \sqrt{2}tr & r^2 & t^2 \\
 \end{pmatrix}
 \end{equation}
\begin{equation}
F_1= 
\begin{pmatrix}
  0 & \sqrt{\frac{1}{2}(1-|t+r|^2)} & \sqrt{\frac{1}{2}(1-|t+r|^2)}  & 0 & 0 & 0 \\
  0 & 0 & 0 & (t+r)\sqrt{\frac{1}{2}(1-|t+r|^2)}  & t\sqrt{\frac{1}{2}(1-|t+r|^2)} & r\sqrt{\frac{1}{2}(1-|t+r|^2)} \\
  0 & 0 & 0 & (t+r)\sqrt{\frac{1}{2}(1-|t+r|^2)} & r\sqrt{\frac{1}{2}(1-|t+r|^2)} & t\sqrt{\frac{1}{2}(1-|t+r|^2)} \\
  0 & 0 & 0 & 0 & 0 & 0 \\
  0 & 0 & 0 & 0 & 0 & 0 \\
  0 & 0 & 0 & 0 & 0 & 0\\
 \end{pmatrix}
 \end{equation}
 \begin{equation}
F_2= 
\begin{pmatrix}
  0 & \sqrt{\frac{1}{2}(1-|t-r|^2)} & -\sqrt{\frac{1}{2}(1-|t-r|^2)}  & 0 & 0 & 0 \\
  0 & 0 & 0 & (r-t)\sqrt{\frac{1}{2}(1-|t-r|^2)}  & t\sqrt{\frac{1}{2}(1-|t-r|^2)} & -r\sqrt{\frac{1}{2}(1-|t-r|^2)} \\
  0 & 0 & 0 & -(r-t)\sqrt{\frac{1}{2}(1-|t-r|^2)} & r\sqrt{\frac{1}{2}(1-|t-r|^2)} & -t\sqrt{\frac{1}{2}(1-|t-r|^2)} \\
  0 & 0 & 0 & 0 & 0 & 0 \\
  0 & 0 & 0 & 0 & 0 & 0 \\
  0 & 0 & 0 & 0 & 0 & 0\\
 \end{pmatrix}
 \end{equation}
     \begin{equation}
F_3= 
\begin{pmatrix}
  0 & 0 & 0 & 0 & \frac{1}{\sqrt{2}}\sqrt{\frac{1}{2}(1-|t-r|^2)}\sqrt{\frac{1}{2}(1-|t+r|^2)}  & -\frac{1}{\sqrt{2}}\sqrt{\frac{1}{2}(1-|t+r|^2)}\sqrt{\frac{1}{2}(1-|t-r|^2)} \\
  0 & 0 & 0 & 0 & 0 & 0 \\
  0 & 0 & 0 & 0 & 0 & 0 \\
  0 & 0 & 0 & 0 & 0 & 0 \\
  0 & 0 & 0 & 0 & 0 & 0 \\
  0 & 0 & 0 & 0 & 0 & 0\\
 \end{pmatrix}
 \end{equation}
      \begin{equation}
F_4= 
\begin{pmatrix}
  0 & 0 & 0 & \frac{1}{\sqrt{2}}(\frac{1}{2}(1-|t+r|^2)) & \frac{1}{2}(\frac{1}{2}(1-|t+r|^2))  & \frac{1}{2}(\frac{1}{2}(1-|t+r|^2)) \\
  0 & 0 & 0 & 0 & 0 & 0 \\
  0 & 0 & 0 & 0 & 0 & 0 \\
  0 & 0 & 0 & 0 & 0 & 0 \\
  0 & 0 & 0 & 0 & 0 & 0 \\
  0 & 0 & 0 & 0 & 0 & 0\\
 \end{pmatrix}
 \end{equation}
       \begin{equation}
F_5= 
\begin{pmatrix}
  0 & 0 & 0 & -\frac{1}{\sqrt{2}}(\frac{1}{2}(1-|t-r|^2)) & \frac{1}{2}(\frac{1}{2}(1-|t-r|^2))  & \frac{1}{2}(\frac{1}{2}(1-|t-r|^2)) \\
  0 & 0 & 0 & 0 & 0 & 0 \\
  0 & 0 & 0 & 0 & 0 & 0 \\
  0 & 0 & 0 & 0 & 0 & 0 \\
  0 & 0 & 0 & 0 & 0 & 0 \\
  0 & 0 & 0 & 0 & 0 & 0\\
 \end{pmatrix}
 \end{equation}
 
\end{widetext}

We can therefore calculate the evolution of the quantum state up to just after the metamaterial by taking the following evolution:
 
 \begin{equation}
 \rho_2 = \sum_i F_i \rho_1 F_i^{\dagger},
 \label{final}
\end{equation}
 
where $\rho_1$ is the state right before the metamaterial (Eq. \ref{rho1}).

\begin{table}
\begin{center}
\begin{tabular}{| c c c |}
    \hline
    {\bf Component} & {\bf Parameter} & {\bf Transmission} \\ 
    \hline
    \hline
   HOM BS & $p_1$ & 0.39 \\ 
     & $q_1$ & 0.19 \\ 
    Intermediate loss & $p_2$ & 0.621 \\ 
     & $q_2$ & 0.626 \\ 
     Output loss & $p_3$ & 0.7 \\ 
     & $q_3$ & 0.6 \\ 
     LBS & $r$ & 0.2991-0.2177i\\
     & $t$ & 0.6625\\
     & $\alpha$ & 0.4758\\
    \hline
 \end{tabular}
 \end{center}
 \caption{Measured transmission coefficients throughout the setup.}
\label{tab1}
\end{table}
 
\begin{figure}[ht!]
\centering{
\includegraphics[width = 0.5\textwidth]{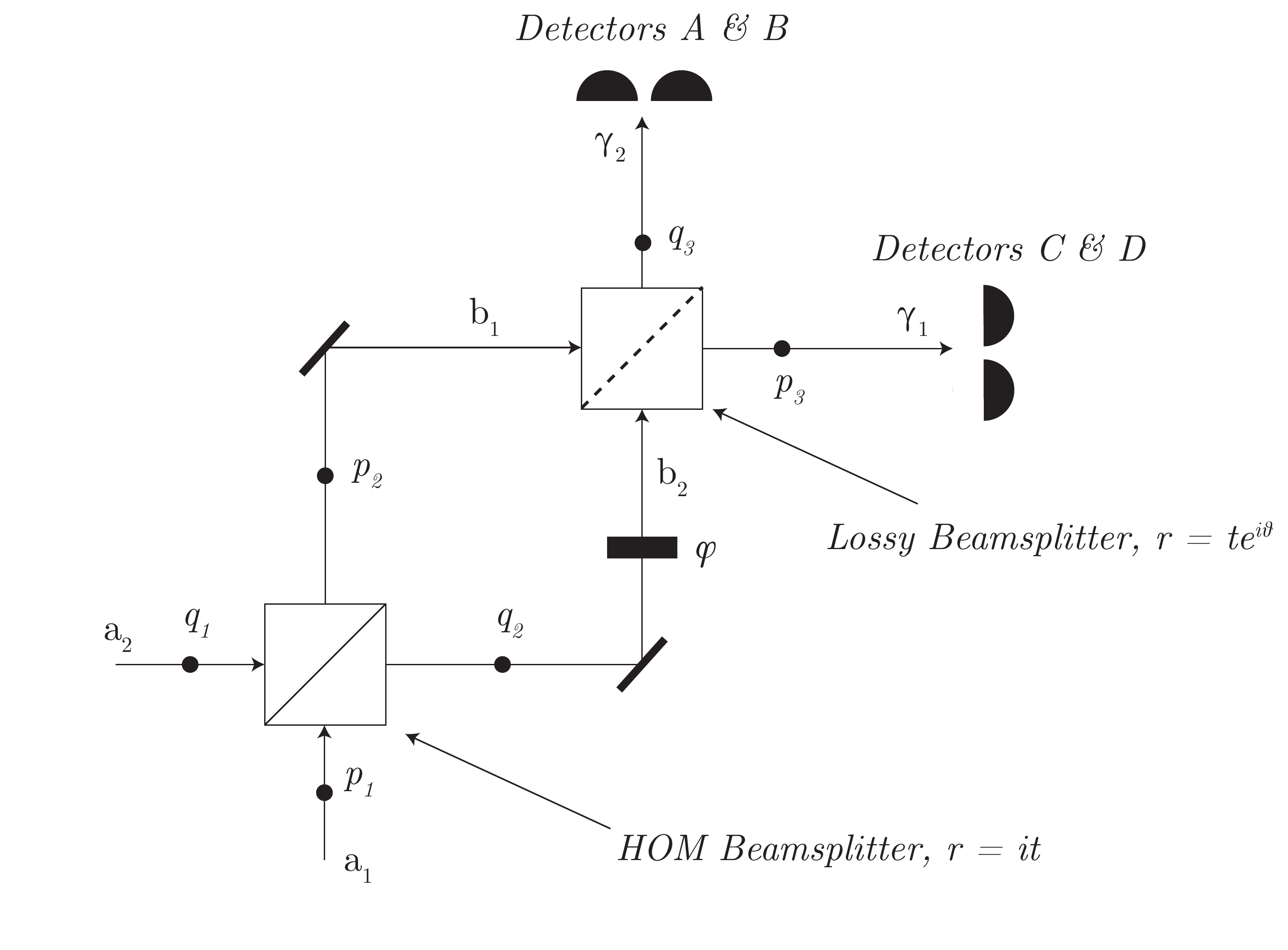}
\caption{Schematic for losses in the coherent absorption of N = 2 N00N state system}
\label{fig:scheme}}
\end{figure}

\begin{widetext}

\subsection{Measurements Calculations}

We assume an input state of $\rho = |11\rangle\langle11|$ from the SPDC source entering the system with parameters shown in Table S2. These values correspond to the scheme set out in the main text. Fig. \ref{fig:scheme} shows a simplified schematic of the system used to study the coherent absorption of N = 2 N00N.\\

Using Eq. \ref{final} we find the equations describing the coincidence counts measured directly after the metamaterial to be,
 
 \begin{align*}
 p(1,1) &= \frac{1}{4}\text{Tr}[|11\rangle \langle 11| \rho_2]\\
           &=\frac{1}{4}(0.00048\sin^2(2\varphi) + 0.00048\cos^2(2\varphi) + 0.00098\cos(2\varphi) + 0.00050),\\
 p(2,0) &= \frac{1}{2}\text{Tr}[|20\rangle \langle 20| \rho_2]\\
           &= \frac{1}{2}(0.00010\sin^2(2\varphi) + 0.00063\sin(2\varphi) + 0.00010\cos^2(2\varphi) +  0.00019\cos(2\varphi) +  0.00109),\\
 p(0,2) &=\frac{1}{2}\text{Tr}[|20\rangle \langle 20| \rho_2]\\
           &= \frac{1}{2}(0.00060\sin^2(2\varphi) -  0.00004\sin(2\varphi) +  0.00060\cos^2(2\varphi) + 0.00001\cos(2\varphi) + 0.00006).
 \end{align*}
\end{widetext}

\section{Maximum Absorption of Two Photons}

We may find the upper bound on the absorption of 2 photons in the coherent absorption scheme, assuming an input state of $|2_+\rangle$ (interferometer phase $\varphi=0$, no trivial losses). Total absorption of two photons takes us to the $N=0$ photons subspace. Therefore, we consider the density matrix element $\rho_{00}$ after the LBS:

\begin{equation}
\rho_{00} = \frac{1}{2}\left[(|r-t|^2-1)^2 + (|r+t|^2-1)^2\right],
\end{equation}

where $r=r_0e^{i\theta}$ and $t$ are the reflection and transmission coefficients of the LBS, and $t,r_0 \in [0,1], \theta\in[0,2\pi)$. By denoting the absorption $\alpha$, which vanishes for a lossless BS,

\begin{align}
\alpha &= 1 - \vert t\vert ^2 - \vert r\vert ^2\\
             &= 1-t^2 - r_0^2,
\end{align}

we obtain the following expression for $\rho_{00}$:

\begin{equation}
\rho_{00} = 4r_0^2(1 - r_0^2 - \alpha)\cos^2\theta + \alpha^2.
\label{rho00}
\end{equation}

Using Eq. 2.4 in \cite{Barnett1998}:

\begin{align}
2tr_0\vert \cos \theta\vert &= 2r_0\sqrt{1-r_0^2 -\alpha}\vert \cos \theta \vert \\
                                      & \leq \alpha.
\end{align}

Therefore, Eq. \ref{rho00} becomes

\begin{align}
\rho_{00} & \leq 2\alpha ^2.
\end{align}

So we find that the element of $|00\rangle\langle00|$ is always less than twice that of the absorption coefficient squared. Therefore given that for a thin film the maximum absorption is $\alpha = 0.5$ we find,

\begin{equation}
\rho_{00} \leq 2 \times 0.5^2 = \frac{1}{2}.
\vspace{0.1cm}
\end{equation}

\section{Metamaterial parameters}
The metamaterial parameters are shown in Fig. 5. Refer to Tab. \ref{tab1}  for experimentally measured reflection and transmission coefficients. The coefficient in experiments is a mixture between parallel and perpendicular polarisation responses as we are using circularly polarised light in order to maximise the throughput of the photons.\\
\begin{figure*}[h!]
\begin{centering}
\includegraphics[width = \textwidth]{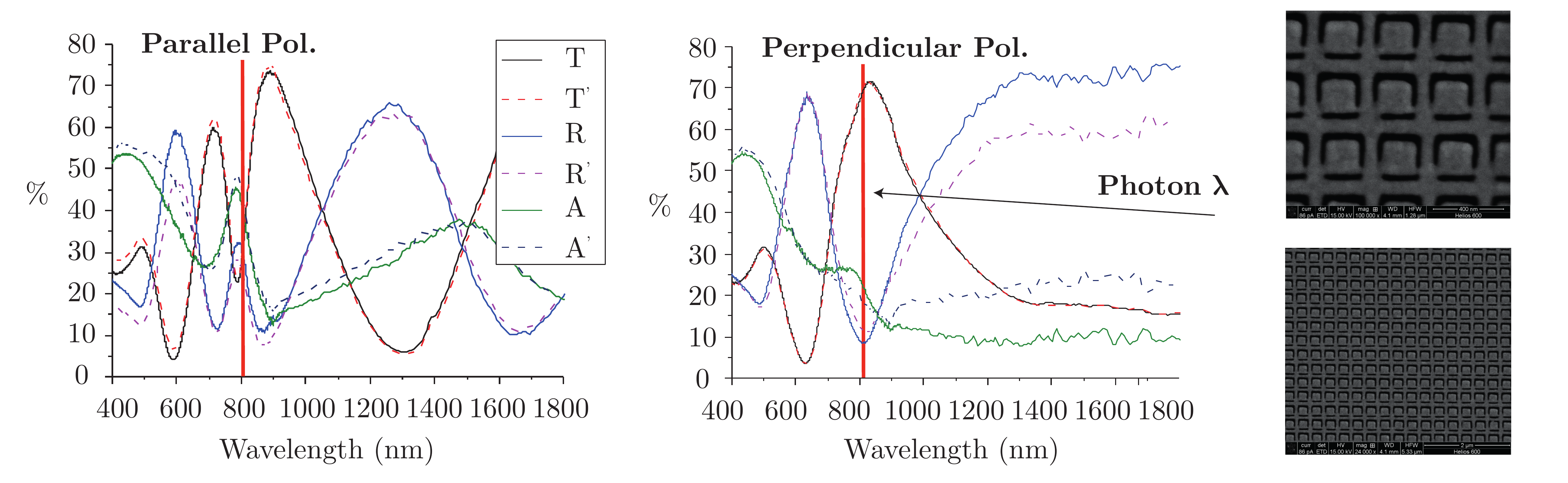}
\caption{Metamaterial parameters measured at Southampton University by Joao Valente for a split-ring resonator array on 50 nm of free-standing gold. The array is milled by a focused ion beam to create a localised plasmon resonance at $\sim$800 nm. The plots show the response for input polarisations parallel (left panel) and perpendicular (middle panel). The reflectance $R$, transmission $T$ and absorption $A$ are shown for input wavelengths between 400 and 1800 nm. The $'$ symbol refers to impinging light to the reverse side of the metamaterial structure.}
\label{fig5}
\end{centering}
\end{figure*}


\bibliographystyle{unsrt}



\end{document}